\documentclass{article}

\usepackage{arxiv}

\usepackage[utf8]{inputenc} 
\usepackage[T1]{fontenc}    
\usepackage{hyperref}       
\usepackage{url}            
\usepackage{booktabs}       
\usepackage{nicefrac}       
\usepackage{microtype}      
\usepackage{lipsum}		
\usepackage{graphicx}
\usepackage{natbib}
\usepackage{doi}

\usepackage{graphicx}
\usepackage{newtxtext}
\usepackage{newtxmath}
\usepackage{natbib}

\newcommand{\RomanNumeralCaps}[1]
\linenumbers

\usepackage{graphicx,color,epsfig,soul,rotating,overpic,varwidth,xcolor}

\usepackage[export]{adjustbox}
\usepackage{subfigure}
\usepackage{subfigmat}
\usepackage{algorithm}
\usepackage[noend]{algpseudocode}
\usepackage{bm}
\usepackage{xfrac}
\usepackage{multicol}
\usepackage{multirow}
\usepackage{pbox}
\usepackage{tabularx}
\usepackage{caption}

\usepackage{booktabs}
\usepackage{tikz}
\usetikzlibrary{tikzmark}
\usetikzlibrary{calc}
\usetikzlibrary{arrows.meta}

\usepackage{lipsum}

\usepackage{placeins}

\usepackage[absolute,overlay,showboxes]{textpos}

\definecolor{blue2}{rgb}{0, 0.4470, 0.7410}
\definecolor{red2}{rgb}{0.8500, 0.1250, 0.0480} 
\definecolor{orange2}{rgb}{0.8500, 0.3250, 0.0980} 
\definecolor{yellow2}{rgb}{0.9290, 0.6940, 0.1250}
\definecolor{purple2}{rgb}{0.4940, 0.1840, 0.5560}
\definecolor{green2}{rgb}{0.4660, 0.6740, 0.1880}
\definecolor{ltblue2}{rgb}{0.3010, 0.7450, 0.9330}
\definecolor{dkred2}{rgb}{0.6350, 0.0780, 0.1840}
\definecolor{gray2}{rgb}{0.22, 0.22, 0.3}
\definecolor{ltgray2}{rgb}{0.647, 0.647, 0.647}
\definecolor{blueIV}{rgb}{0, 0, 0.7410}
\definecolor{blueIII}{rgb}{0.2, 0.2, 0.7410}
\definecolor{blueII}{rgb}{0.4, 0.4, 0.7410}
\definecolor{blueI}{rgb}{0.7410, 0.7410, 0.7410}
\definecolor{jetVI}{rgb}{0.9763 0.9831 0.0538}
\definecolor{jetV}{rgb}{0.9264 0.7256 0.2996}
\definecolor{jetIV}{rgb}{0.4783 0.7489 0.4877}
\definecolor{jetIII}{rgb}{0.0282 0.6663 0.7574}
\definecolor{jetII}{rgb}{0.0582 0.4677 0.8589}
\definecolor{jetI}{rgb}{0.2081 0.1663 0.5292}
\definecolor{mode1}{rgb}{0 0.4470 0.7410}
\definecolor{mode2}{rgb}{0.8500 0.3250 0.0980}
\definecolor{mode3}{rgb}{0.9290 0.6940 0.1250}
\definecolor{dkgold}{rgb}{0.5930 0.5150 0.3260}
\definecolor{mode1}{rgb}{0 0.4470 0.7410}
\definecolor{mode2}{rgb}{0.8500 0.3250 0.0980}
\definecolor{mode3}{rgb}{0.9290 0.6940 0.1250}
\definecolor{dkgold}{rgb}{0.5930 0.5150 0.3260}
\definecolor{matblue}{rgb}{0.000 0.447 0.741}
\definecolor{matred}{rgb}{0.850 0.325 0.098}
\definecolor{matyellow}{rgb}{0.9290 0.6940 0.125}
\definecolor{matpurple}{rgb}{0.494 0.184 0.556}
\definecolor{matgreen}{rgb}{0.466 0.674 0.188}
\definecolor{matcyan}{rgb}{0.3010 0.7450 0.9330}

\newcommand{\etal}{\textit{et al.}}

\usepackage{soul}

\usepackage{hyperref}
\hypersetup{
	colorlinks = true,
	urlcolor = blue,
	citecolor = blue,
}

\title{A sensor-restrained artificial shear diffusivity for large-eddy simulations of vortex-dominated compressible flows}

\author{
	{Jean H\'elder Marques Ribeiro\thanks{Currently at Universidade Federal Fluminense, email: jeanmarques@id.uff.br}, Hugo Felippe da Silva Lui and William Roberto Wolf} \\
	Faculdade de Engenharia Mec\^{a}nica\\ 
	Universidade Estadual de Campinas \\
	13083-860 Campinas, SP, Brazil\\
}

\renewcommand{\shorttitle}{Sensor-restrained artificial shear diffusivity for LES}

\begin{document}

\maketitle
\begin{abstract}
    We propose a sensor-restrained model for the shear viscosity term within the localized artificial diffusivity (LAD) scheme to stabilize compressible large-eddy simulations with {low-pressure-core} vortical structures. {LAD methods are used in numerical solvers  based on spectral-like compact finite-difference schemes. While high-order-accurate numerical schemes with proper discretization guarantees physical fidelity, the LAD role is to suppress non-physical oscillations arising in compressible flow simulations near shock waves and other sharp gradients. LAD is a cost-effective approach which adds artificial shear and bulk viscosities, and thermal conductivity to their physical counterparts. However, an unrestricted added diffusivity may lead to poorly-resolved coherent structures and undesirable turbulence statistics.} In order to prevent excessive numerical diffusion in compressible shear flows, the artificial shear viscosity term {can be disabled in cases where the simulation is already stable}. However, in flow simulations where vortices emerge within a low-pressure region, a strong pressure decay may lead to instabilities that make the simulation unstable. For such cases, adding artificial shear viscosity is necessary to maintain numerical stability, as this issue is unaddressed by the artificial bulk viscosity and thermal conductivity alone. Our approach integrates a sensor into the standard LAD formulation, particularly in the artificial shear viscosity, that reduces the added diffusivity while preserving numerical stability. This advancement is possible by adding the shear diffusivity only in localized flow regions consisting of low-pressure-core vortices, and it enables {stable and accurate large-eddy simulations (LES) of compressible vortex-dominated flows, such as those encountered in separated flows and bluff-body wakes}.  
\end{abstract}

\section{Motivation}
\label{sec:intro}
\label{sec:intro}
In the context of transitional and turbulent compressible flows, large eddy simulations (LES) play a crucial role in revealing the fundamental unsteady flow features. Through accurate numerical schemes, it is possible to investigate the details of processes undergoing multiple scale phenomena such as fluid mixing, sound generation, transition to turbulence, shock-boundary layer interactions, to name a few. For high-Reynolds-number flows, the numerical approach must be able to accurately resolve the broad range of spatial and temporal scales associated with turbulence. Accurate solutions of turbulent flows can often be achieved through high-order compact finite-difference schemes with spectral-like resolution \cite{Lele:JCP92}. These methods are known to reduce numerical errors associated with dispersion and dissipation. Albeit being applied to many practical applications \cite{VisbalGaitonde:AIAAJ99}, in the presence of steep gradients and discontinuities, non-physical numerical oscillations can be generated making the simulation unstable. Developing numerical solvers able to resolve turbulence and cope with discontinuities in the flow remains a challenging task.

An elegant solution to this problem was initially proposed by Cook and Cabot \cite{Cook:JCP2004high}, later being refined by Cook \cite{Cook:PF2007artificial}. This approach, known as localized artificial diffusivity (LAD), offers several benefits, including low computational cost, simple implementation, and automatic deactivation in smooth regions. The core idea is to introduce artificial fluid properties that suppress unresolved high-frequency content in the flow, thereby preventing spurious oscillations. In its original formulation \cite{Cook:PF2007artificial}, the LAD successfully stabilized simulations without significant alterations on turbulence statistics. Due to its effectiveness, LAD schemes have been employed in a diverse range of physical problems \cite{Wang:PRF2021thermal,Lui:PRF22,Cherng:PRF2024heat}. {Understandably, an ideal LAD model would be able to suppress non-physical oscillations while minimizing the introduction of artificial diffusivity properties, hence achieving both numerical accuracy and stability.} For this reason, LAD developments involve spatially restraining the artificial fluid properties to specific regions of interest using appropriate sensors \cite{Jameson1981numerical,Ducros:JCP1999sensor}. 

Kawai \etal~\cite{Kawai:JCP2010assessment} further advanced LAD models by focusing on the computation of artificial shear and bulk viscosities, and thermal conductivity, along with a detailed assessment of sensors for compressible flows. {Inspired by the physical aspects of the flow near the source of numerical instabilities,} the authors applied a combination of a Ducros-type sensor \cite{Ducros:JCP1999sensor} with a Heaviside function of negative dilatation to the artificial bulk viscosity. This approach effectively minimized unnecessary dissipation far from compression waves. Similar sensors have been widely adopted in compressible flow simulations to identify shock waves, even beyond the LAD context \cite{KhalloufiCapecelatro:IJMF2023drag,TamakiAIAAJ24transonic}. {Despite of its effectiveness for spatially sparsifying the application of bulk viscosity, such sensors have never been applied to the artificial shear viscosity. In fact, Olson \& Lele \cite{Olson:JCP2013directional} recommended turning off this term if the numerical scheme is already stable. These authors showed that artificial shear viscosity can compromise flow statistics when applied excessively along turbulent boundary layers.}

While LAD without artificial shear viscosity has been effective near shock waves, we note that numerical simulations of supersonic flows may still experience instabilities when strong vortices emerge in low-pressure expansion regions. {As the bulk viscosity is ineffective in these zones, turning on the artificial shear viscosity is required.} To balance the need for resolution in the turbulent scales with numerical stability, we propose a modification in the application of the artificial shear viscosity using a Ducros-type sensor, inspired by the methodology from Kawai \etal~\cite{Kawai:JCP2010assessment}. Our approach spatially restrains the application of artificial shear viscosity, while maintaining numerical stability. The sensor is constructed to avoid high-shear regions, such as boundary layers and wakes. Our work is organized as follows: Sections \ref{sec:math_governing_equations} and \ref{sec:math_lad}  present the governing equations, the details of our numerical solver, and the standard LAD. Section \ref{sec:math_lad_modified} introduces our sensor modification to the artificial shear viscosity. Section \ref{sec:results} discusses the results of 2D test cases and a 3D turbulent compressible flow with strong vortices that arise in a region of expansion, and Section \ref{sec:conclusions} presents our conclusions.

\section{Theoretical and numerical formulation}
\label{sec:math}

\subsection{Governing equations and numerical schemes}
\label{sec:math_governing_equations}

The compressible Navier-Stokes equations for a calorically perfect gas can be written as


\begin{equation}
\frac{\partial \rho}{\partial t} + \nabla \cdot \left( \rho \boldsymbol{u} \right) = 0 \mbox{ ,}
\label{eq:continuity}
\end{equation}


\begin{equation}
\frac{\partial \left( \rho  \boldsymbol{u} \right)}{\partial t} + \nabla \cdot \left[\rho \boldsymbol{u} \boldsymbol{u} + p \underline{\delta} - \underline{\tau} \right]
=  0 \mbox{ ,}
\end{equation}
and
\begin{equation}
\frac{\partial E}{\partial t}
+ \nabla \cdot \left[E \boldsymbol{u}  + \left(p \underline{\delta} -  \underline{\tau}\right) \cdot \boldsymbol{u}  - \kappa \nabla T  \right] = 0 \mbox{ ,} 
\end{equation}
where
\begin{equation}
E = \frac{p}{\gamma - 1} + \frac{1}{2} \rho \boldsymbol{u} \cdot \boldsymbol{u} \mbox{, } \hspace{8pt} \underline{\tau} = \mu \left(2 \underline{S} \right) + \left(\beta - \frac{2}{3}\mu \right) \left(\nabla \cdot \boldsymbol{u}\right) \underline{\delta} \mbox{ ,}
\label{eq:total_energy}
\end{equation}
and $p = \rho R T $ is the equation of state. Here, $\rho$ is the fluid density, $\boldsymbol{u}$ is the velocity vector, $\underline{\delta}$ is the Kronecker tensor, $T$ is the static temperature, $p$ is the pressure,  $\beta$ is the bulk viscosity, $\underline{S}$ is the strain rate tensor, $\mu$ is the dynamic viscosity, related to $T$ through Sutherland's law (Sutherland constant set to $S_{\mu}/T_{\infty}$ = 0.368), and $\kappa$ is the thermal conductivity computed using the Prandtl number $Pr = {c_p \mu}/{\kappa} $, where $c_p$ is the specific heat at constant pressure.

In the present work, the governing equations are solved in non-dimensional form using a sixth-order accurate compact finite-difference scheme for the spatial discretization~\cite{Nagarajan:JPC03}
{and advanced in time separately for each zone of the domain. Near the wall, an implicit modified second-order Beam–Warming scheme is employed, while an explicit third-order low-storage Runge-Kutta scheme is applied far from the wall. 
The zones are connected by an overlap region and a fourth-order Hermite interpolation is used to exchange data between zones \cite{Bhaskaran:10}. High-wavenumber numerical instabilities originated from mesh non-uniformities and interpolations are suppressed at each time step by a sixth-order compact filter \cite{Lele:JCP92}. We set no-slip adiabatic boundary conditions at the wall and characteristic boundary conditions based on Riemann invariants and sponge layers at the farfield. For three-dimensional (3D) flows, we apply periodic boundary conditions in the spanwise direction. Further details of the numerical methodology can be found in the Refs.~\cite{Nagarajan:JPC03,Bhaskaran:10}.}


\subsection{Standard LAD formulation}
\label{sec:math_lad}

In its standard form~\cite{Cook:PF2007artificial,Kawai:JCP2010assessment}, we compute local values of artificial shear viscosity $\mu^*$, bulk viscosity $\beta^*$, and thermal conductivity $\kappa^*$ to be added to the fluid properties as
$\mu = \mu_f + \mu^*$, $\beta = \beta_f + \beta^*$, and $\kappa = \kappa_f + \kappa^*$, where $f$ and $^*$ are used for fluid and LAD properties, respectively. The generalized LAD scheme is defined as
\begin{eqnarray}
\mu^* &=& C_\mu \overline{\rho \left| \Sigma_{l = 1}^3 \frac{\partial^r \mathcal{F_\mu}}{\partial \xi_l^r} \Delta \xi_{l}^{r} \Delta_{l,\mu}^2\right|} \mbox{  ,} \label{eq:LADshear}\\
\beta^* &=& C_\beta \overline{\rho f_{sw} \left| \Sigma_{l = 1}^3 \frac{\partial^r \mathcal{F_\beta}}{\partial \xi_l^r} \Delta \xi_{l}^{r}  \Delta_{l,\beta}^2\right|} \mbox{  ,}\\
\kappa^* &=& C_\kappa \overline{\frac{\rho c_s}{T} \left| \Sigma_{l = 1}^3 \frac{\partial^r \mathcal{F_\kappa}}{\partial \xi_l^r} \Delta \xi_{l}^{r} \Delta_{l,\kappa}^2\right|} \mbox{  ,}
\label{eq:LADmodel}
\end{eqnarray}
where $C_{\mu}$, $C_{\beta}$, and $C_{\kappa}$ are user-specified constants, defined as $C_{\beta} = 1.75$ and $C_{\kappa} = 0.01$ throughout this work. The constant $C_\mu$ will be part of our later discussion. The term $c_s$ is the local speed of sound, and $\Delta \xi_{l}$ and $\Delta_{l,\cdot}$ are the grid spacing in the computational space and physical space, respectively~\cite{Kawai:JCP2010assessment}. The overbar denotes an approximate truncated Gaussian filter \citep{Cook:PF2007artificial}. {The functions $\mathcal{F_\mu}$, $\mathcal{F_\beta}$, and $\mathcal{F_\kappa}$ are designed to detect unresolved eddies. These functions appear within a polyharmonic operator, which denotes a series of Laplacian operators. Following Cook~\cite{Cook:PF2007artificial}, we set $r = 4$, which denotes a biharmonic operator able to damp high-wavenumber oscillations, similar to a spectral vanishing viscosity \cite{Karamanos:JCP2000spectral}}. The functions $\mathcal{F_\beta}$, $\mathcal{F_\mu}$ and $\mathcal{F_\kappa}$, and the shock sensor for $\beta^*$, $f_{sw}$, are set in accordance with the LAD-D2-0 scheme from Kawai \etal~\cite{Kawai:JCP2010assessment}.

{The three artificial properties of the LAD scheme, $\mu^*$, $\beta^*$, and $\kappa^*$, have been employed within non-dissipative compact-finite-differencing schemes for high-fidelity numerical simulations of turbulent compressible flows in different scenarios \cite{Kawai:JCP2008localized,Kawai:AIAAJ2010les,Wang:PRF2021thermal,Lui:PRF22,Cherng:PRF2024heat,Lui:TCFD24mach}. For several simulations involving shock waves, the sole application of $\beta^*$ and $\kappa^*$ is sufficient to ensure numerical stability. In such cases, Olson \& Lele \cite{Olson:JCP2013directional} recommend to turn off $\mu^*$, as its unrestrained application over turbulent boundary layers substantially affects the quality of the flow statistics. Moreover, when added without restriction to high-shear zones, $\mu^*$ may act as an external flow disturbance in a region known to be susceptible to spatial-temporal amplification  \cite{Madhusudanan:JFM2025resolvent,Ribeiro:JFM2024control}, leading to further spurious oscillations.}

\subsection{A sensor-restrained model for artificial shear viscosity $\mu^*$}
\label{sec:math_lad_modified}
{Numerical stability is attained by the standard LAD. It is a paramount goal to maintain this capability while minimizing the addition of $\mu^*$ over high-shear zones such as boundary layers and wakes. In this work, we propose a sensor $f_{sh}$ for $\mu^*$ to restrict its application to low-pressure-core regions where vortices cause extreme pressure drops, making the simulation unstable.} Our formulation modifies the equation for $\mu^*$ as follows:
\begin{equation}
\mu^* = C_\mu \overline{\rho f_{sh} \left| \Sigma_{l = 1}^3 \frac{\partial^4 \mathcal{F_\mu}}{\partial \xi_l^4} \Delta \xi_{l}^{4} \Delta_{l,\mu}^2\right|} \mbox{ ,}
\label{eq:muArt}
\end{equation}
where 
{
\begin{equation}
	f_{sh} =
	\underbrace{H(Q)}_{\text{Q-based term}} 
	\quad
	\underbrace{H(p - p_\text{set})}_{\text{pressure-based term}}
	\quad 
	\underbrace{\frac{ |\nabla \times \boldsymbol{u}|^2}{ \left(\nabla \cdot \boldsymbol{u}\right)^2 +  |\nabla \times \boldsymbol{u}|^2+ \epsilon}}_{\text{vorticity-dilatation-based term}}
	\label{eq:f_sh3D}
\end{equation}}is a real-valued $\mu^*$ sensor ($f_{sh} \in [0,1]$). Here, $H(\cdot)$ is the Heaviside function and {$\epsilon = 10^{-16}$ is a predefined small positive real number} chosen to prevent numerical issues in regions where both divergence of velocity $\left(\nabla \cdot \boldsymbol{u}\right)$ and vorticity $\left(\nabla \times \boldsymbol{u}\right)$ are null \cite{Ducros:JCP1999sensor,Kawai:JCP2010assessment}. The sensor $f_{sh}$  is composed of three main parameters: a $Q$-criterion-based, a pressure-based, and a vorticity-dilatation-based term. 

For the first term, $Q$ is the second invariant of the velocity gradient tensor, used to identify vortices when $Q > 0$ \cite{Hunt:CTR1988eddies}. The term $p_\text{set}$ denotes a threshold of $p$. The sensor activates $\mu^*$  for critical regions where the local pressure drops below $p_\text{set}$. The vorticity-dilatation-based term, similar to that proposed by Kawai \etal~\cite{Kawai:JCP2010assessment}, maintains the denominator structure and places the vorticity magnitude in the numerator, such that $\mu^*$ is activated in shearing regions untackled by $\beta^*$. In the following section, we explore the application of the sensor-restrained $\mu^*$ and compare it to the standard $\mu^*$ in various flow problems. 


\section{Numerical results}
\label{sec:results}

\subsection{2D test cases of transonic flows}
\label{sec:2d_transonic}
{To test the capability of the sensor-restrained $\mu^*$ in adding minimal but sufficient diffusivity to stabilize numerical simulations, we study flows over NACA0012 airfoils with static stall at angle of attack $\alpha = 8^\circ$ and undergoing dynamic stall through pitching motion (see Ref. \citep{Miotto:AIAAJ22} for kinematic details). For both cases, the chord-based Reynolds number is set to $Re_c = 200,000$ and the freestream Mach number $M_\infty = 0.7$, with the fluid modeled as a calorically perfect gas with $\gamma = 1.4$ and $Pr = 0.7$.}

{The Cartesian coordinate system is centered at the leading edge of the airfoil, $(x, y)/c = (0,0)$, and the computational domain spans $(x, y)/c \in [-20, 20] \times [-20, 20]$. A body-fitted O-grid is employed, consisting of $n_\xi = 864$ points along the airfoil surface and $n_\eta = 600$ points in the wall-normal direction. The wall-normal grid is divided into $407$ points for the near-wall implicit time-marching scheme and $208$ points for the third-order explicit scheme, with a $15$-point overlap region. The wall-normal spacing is limited to $\Delta\eta_{\text{min}}/c = 0.0005$. The wall-tangential spacing, $\Delta\xi$, is smoothly distributed along the airfoil, being limited to $\Delta\xi_{\text{min}} = 2\Delta\eta_{\text{min}}$ near the leading and trailing edges.} 

\begin{figure}[!t]
\footnotesize
\centering
\begin{tikzpicture}
\node[anchor=south west,inner sep=0] (image) at (0,0) {\includegraphics[page=1,trim=0mm 0mm 0mm 0mm, clip,width=1\textwidth]{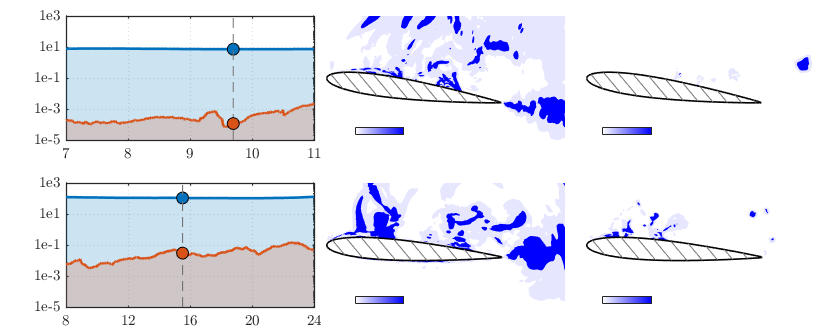}};
\node[anchor=east] at (0.80,6.60) {$(a)$};
\node[anchor=east] at (7.00,6.60) {$(b)$};
\node[anchor=east] at (12.00,6.60) {$(c)$};
\node[anchor=east] at (0.80,3.20) {$(d)$};
\node[anchor=east] at (7.00,3.20) {$(e)$};
\node[anchor=east] at (12.00,3.20) {$(f)$};
\node[anchor=east] at (9.90,6.60) {\color{matblue}standard $\mu^*$};
\node[anchor=east] at (15.10,6.60) {\color{matred}sensor-restrained $\mu^*$};
\node[anchor=east] at (0.60,5.05) {$\phi$};
\node[anchor=east] at (0.60,1.70) {$\phi$};
\node[anchor=east] at (4.40,-0.10) {$\alpha(^\circ)$};
\node[anchor=east] at (4.00,3.20) {$t$};
\scriptsize
\node[anchor=east] at (8.05,4.25) {$\mu^*/\mu$};
\node[anchor=east] at (7.50,3.70) {$1\mathrm{e}{-4}$};
\node[anchor=east] at (8.65,3.70) {$1\mathrm{e}{-2}$};
\node[anchor=east] at (13.00,4.25) {$\mu^*/\mu$};
\node[anchor=east] at (12.45,3.70) {$1\mathrm{e}{-4}$};
\node[anchor=east] at (13.60,3.70) {$1\mathrm{e}{-2}$};
\node[anchor=east] at (8.05,0.90) {$\mu^*/\mu$};
\node[anchor=east] at (7.50,0.35) {$1\mathrm{e}{-2}$};
\node[anchor=east] at (8.45,0.35) {$1\mathrm{e}{0}$};
\node[anchor=east] at (13.00,0.90) {$\mu^*/\mu$};
\node[anchor=east] at (12.45,0.35) {$1\mathrm{e}{-2}$};
\node[anchor=east] at (13.40,0.35) {$1\mathrm{e}{0}$};
\end{tikzpicture} \\ \vspace{-4mm}
\caption{Comparison between standard and sensor-restrained $\mu^*$ . Results for $(a$-$c)$ static and $(d$-$f)$ pitching airfoils are shown. Colored circles mark the time and angle-of-attack instants used for visualizations $(b,c,e,f)$.} 
\label{fig:muArt_over_time}
\end{figure}
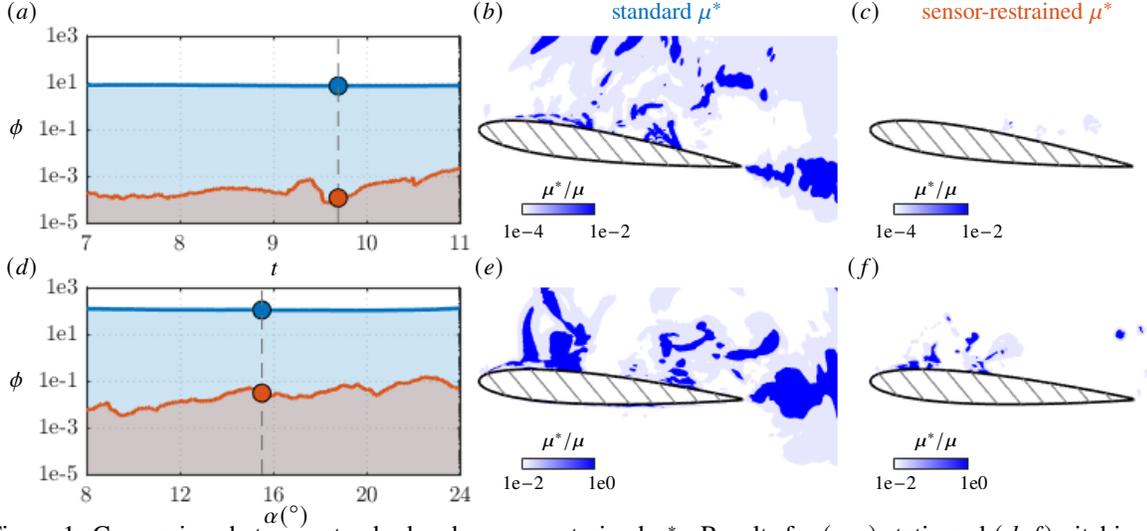

As our goal is to reduce the added artificial diffusivity while maintaining stability, we present the spatially integrated $\mu^*$ in Fig.~\ref{fig:muArt_over_time} for different instants of $(a)$ time and $(d)$ angle of attack. The top row presents results for the static airfoil and the bottom row shows results for the dynamic stall case. For both cases, $p_\text{set} = p_\infty / 4$. Note that the sensor-restrained $\mu^*$ achieves two key objectives of the LAD: (i) ensuring numerical stability \cite{Cook:PF2007artificial,Kawai:JCP2010assessment} and, once stability is attained, (ii) minimizing the addition of artificial diffusivity \cite{Olson:JCP2013directional} over extended time periods for both 2D cases. 

{The quantity $\phi = \int_{V} (\mu^* / \mu) \rm dv$ is introduced as the volume integral of $\mu^* / \mu$ over the computational domain $V$ at different time instants. By analyzing $\phi$ over time, we show that the artificial diffusivity introduced by $\mu^*$ is 3 to 4 orders of magnitude lower for the sensor-restrained $\mu^*$, when compared to the standard LAD. Furthermore, as seen in the lower panel of Fig. \ref{fig:muArt_over_time} for the dynamic stall case, the addition of $\mu^*$ increases with $\alpha$ as the flow separation becomes more pronounced. In fact, the dynamic stall test case generates stronger vortices, requiring $C_\mu = 0.4$, while $C_\mu = 0.002$ proved sufficient for the static airfoil.  We note that the present 2D flows are model problems and their physical aspects differ from turbulent flows. For this reason, they are shown here as test cases for our methodology. Notwithstanding, the insights obtained from these simple 2D cases are closely related to the observations in more complex 3D turbulent compressible flows. These flows are explored in the next section, in which we extend our methodology beyond 2D test cases to a 3D turbulent supersonic flow over a curved blade}.

\subsection{3D supersonic turbulent flow}
\label{sec:2d_transonic}
To attest our methodology for compressible turbulent flows, we employ the sensor-restrained $\mu^*$ to maintain numerical stability for a 3D supersonic turbulent flow over a blade with curved surface. The freestream Mach number is $M_\infty = 2$ and the chord-based Reynolds number is $Re_c = 500,000$. {The fluid is a calorically perfect gas with  $\gamma = 1.4$ and $Pr = 0.72$. The computational domain spans $(x, y, z)/c \in [-2, 4] \times [-2, 2]  \times [0, 0.16]$.}

{In this simulation, an overset mesh is employed with two overlapping grids. The first is a body-fitted O-grid block ($1681 \times 280 \times 252$), which is used to accurately resolve the turbulent boundary layers around the blade, while the second is an H-grid block ($960 \times 491 \times 126$). Overall, the computational mesh consists of approximately $178 \times 10^{6}$ points. The grid resolution follows the standard guidelines for wall-resolved LES \cite{Georgiadis2010}. The near-wall grid spacing, in terms of wall units, is limited to $\Delta_s^{+} \approx 27$, $\Delta_n^{+} \approx 0.8$, and $\Delta_z^{+} \approx 10$, where the subscripts $s$, $n$, and $z$ represent the streamwise, wall-normal and spanwise flow coordinates, respectively}. 

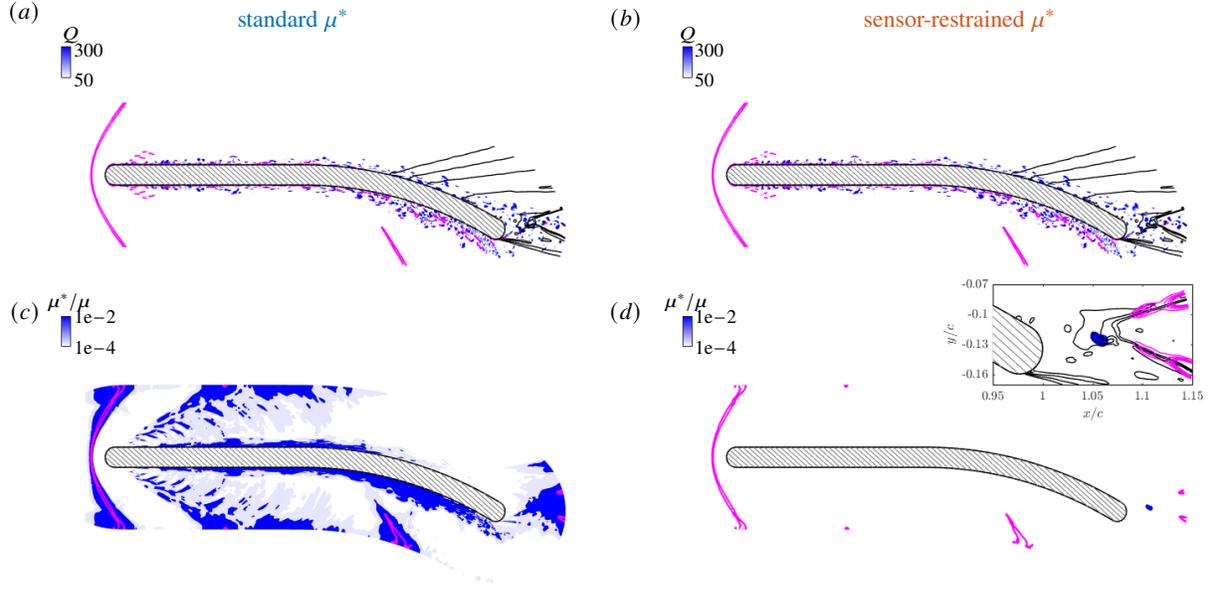
\begin{figure}[!t]
\footnotesize
\centering
\begin{tikzpicture}
\node[anchor=south west,inner sep=0] (image) at (0,0) {\includegraphics[page=1,trim=0mm 0mm 0mm 0mm, clip,width=1\textwidth]{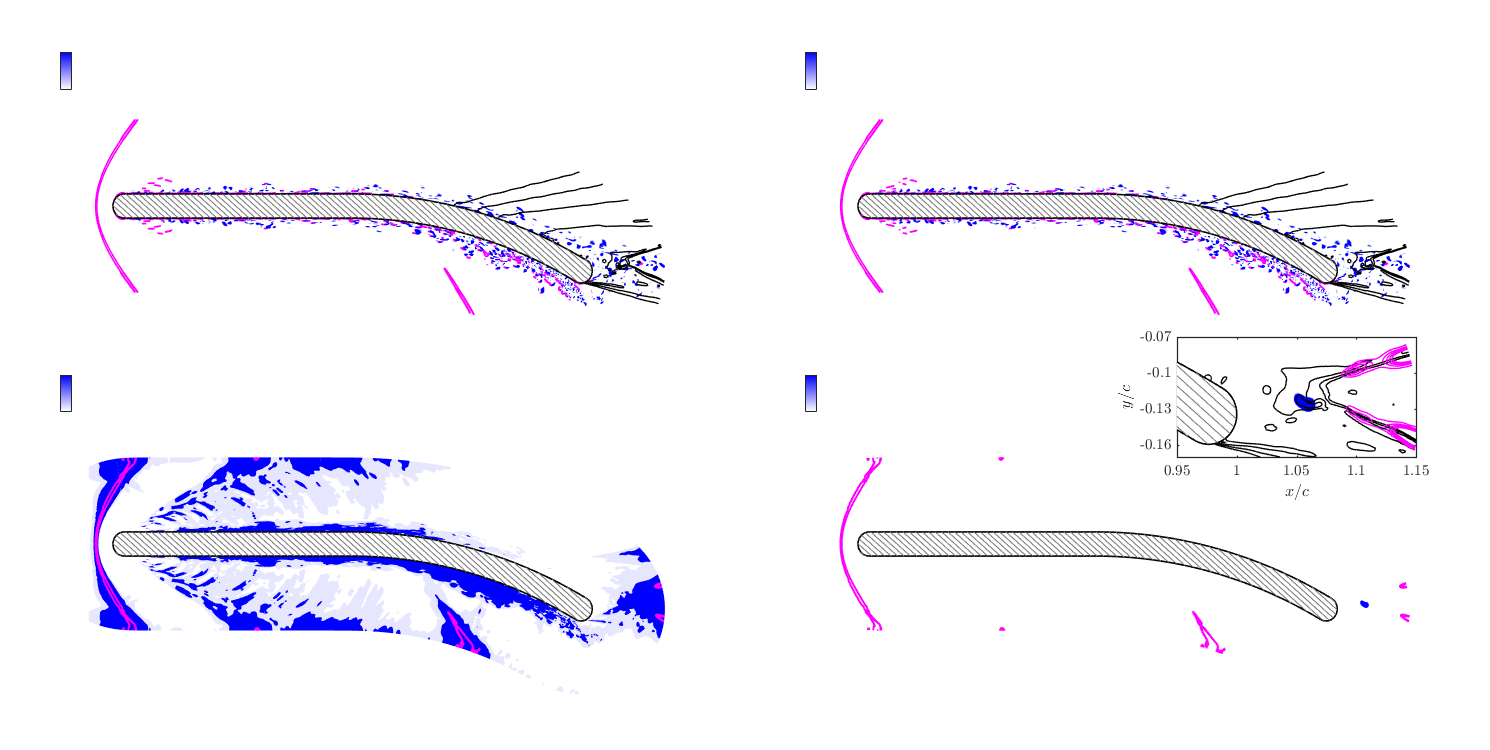}};
\node[anchor=east] at (0.50,8.20) {$(a)$};
\node[anchor=east] at (4.60,8.10) {\color{matblue}standard $\mu^*$};
\node[anchor=east] at (8.50,8.10) {$(b)$};
\node[anchor=east] at (14.00,8.10) {\color{matred}sensor-restrained $\mu^*$};
\node[anchor=east] at (0.50,4.20) {$(c)$};
\node[anchor=east] at (8.50,4.20) {$(d)$};
\scriptsize
\node[anchor=east] at (1.00,7.90) {$Q$};
\node[anchor=west] at (0.75,7.70) {$300$};
\node[anchor=west] at (0.75,7.30) {$50$};
\node[anchor=east] at (1.15,4.35) {$\mu^*/\mu$};
\node[anchor=west] at (0.75,4.15) {$1\mathrm{e}{-2}$};
\node[anchor=west] at (0.75,3.75) {$1\mathrm{e}{-4}$};
\node[anchor=east] at (9.20,7.90) {$Q$};
\node[anchor=west] at (9.00,7.70) {$300$};
\node[anchor=west] at (9.00,7.30) {$50$};
\node[anchor=east] at (9.35,4.35) {$\mu^*/\mu$};
\node[anchor=west] at (9.00,4.15) {$1\mathrm{e}{-2}$};
\node[anchor=west] at (9.00,3.75) {$1\mathrm{e}{-4}$};
\end{tikzpicture} \\ \vspace{0mm}
\caption{Mid-span visualizatons of sensor-restrained $\mu^*$ for the supersonic flow over a blade at $M_\infty = 2$, $Re_c = 500,000$ {at the same instant in time}. $(a,b)$ Black isolines of $p = [0,0.35]$ and magenta isolines of $\| \nabla \rho \| = [100,150]$. $(c,d)$ Magenta isolines of $\beta^*/\mu = [10,50]$. THe insert shows a zoomed-in view with contours of $\mu^*/\mu$, magenta isolines of $\beta^*/\mu$, and black isolines of $p$.} 
\label{fig:Hugo}
\end{figure}

On the convex surface, the combination of vorticity-related pressure decay and an expansion-related low pressure region may lead to a locally steep gradient with negative pressure values, making the simulation unstable without $\mu^*$. The standard $\mu^*$ stabilizes the simulation while introducing artificial dissipation over large portions of the domain. These include the shock wave, which is already covered by $\beta^*$. A proper sensor-restrained $\mu^*$ achieves numerical stability and also avoids adding $\mu^*$ over the boundary layers and high-shear zones, as shown in Fig. \ref{fig:Hugo}. In the left-column, Figs. \ref{fig:Hugo}$(a,c)$, the standard $\mu^*$ is employed. In the right-column, Figs. \ref{fig:Hugo}$(b,d)$, the sensor-restrained $\mu^*$ is used with $p_\text{set} = p_\infty / 4$. The choice of $p_\text{set}$ guarantees that only low-pressure-core vortices evolving far from the boundary layer and the blade are dissipated by the LAD method.
An inset in the top corner of Fig. \ref{fig:Hugo}$(d)$ zoom in on the region $(x,y)/c \in [0.95,1.15] \times [-0.17,-0.07]$ and showcases the small blue region where $\mu^*$ is applied to stabilize the simulation.

For this supersonic flow, we introduce $\mu^*$ with $C_\mu = 0.4$ shortly before the simulation becomes unstable. In spite of the large value for $C_\mu$, the artificial diffusivity $\mu^*$ is substantially lower than the physical viscosity $\mu$. While $\beta^*$ is enforced over bulk compression regions, the sensor-restrained $\mu^*$ is applied in regions not addressed by $\beta^*$ mainly due to the vorticity-dilatation term of the sensor. The sensor $f_{sh}$ restricts $\mu^*$ to the low-pressure region over the curved geometry, as shown in Fig. \ref{fig:Hugo}$(d)$. Concatenating the $Q$-based and $p$-based terms within $f_{sh}$, further constrains $\mu^*$ to a small region within the wake, avoiding the addition of unnecessary diffusivity, while maintaining numerical stability.  

\begin{figure}[!t]
\footnotesize
\centering
\begin{tikzpicture}
\node[anchor=south west,inner sep=0] (image) at (0,0.0) {\includegraphics[page=1,trim=0mm 0mm 0mm 0mm, clip,width=1\textwidth]{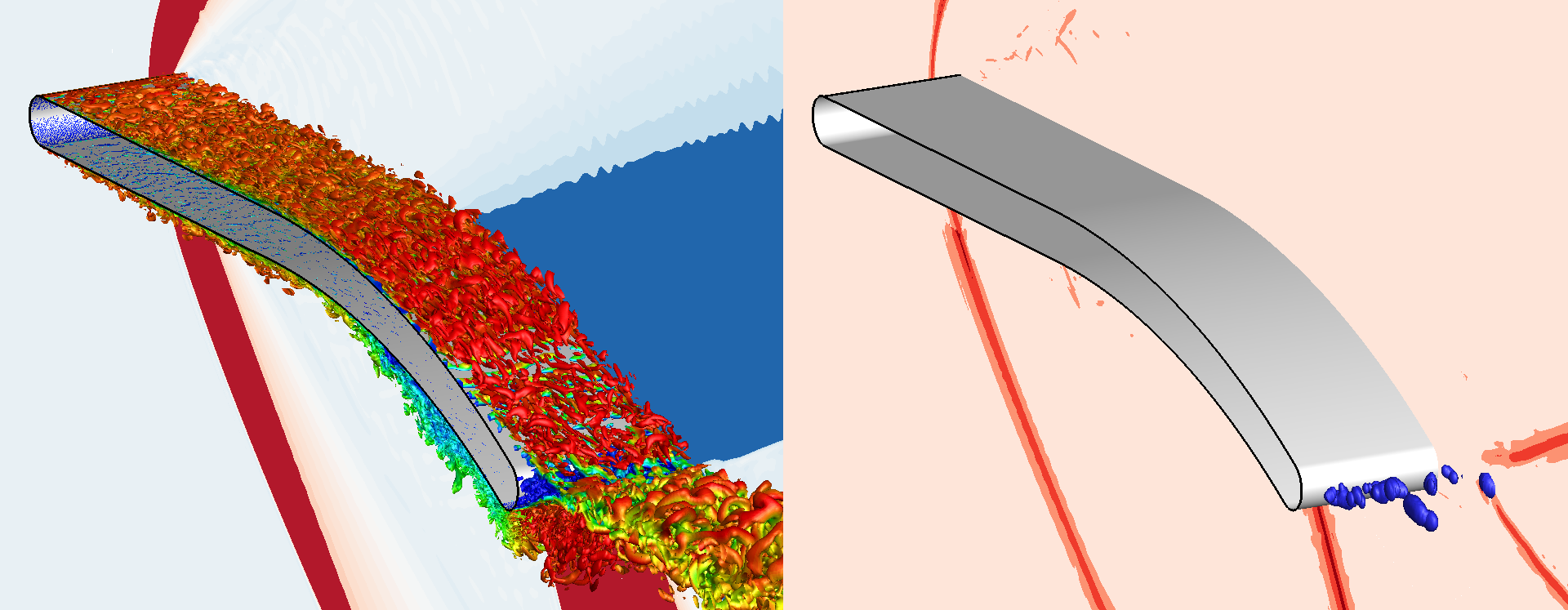}};
\node[anchor=south west,inner sep=0] (image) at (0.08,1.20) {\includegraphics[page=1,trim=35mm 0mm 28mm 0mm, clip,width=0.018\textwidth]{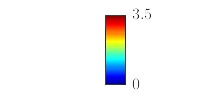}};
\node[anchor=south west,inner sep=0] (image) at (0.10,0.02) {\includegraphics[page=1,trim=35mm 0mm 28mm 0mm, clip,width=0.013\textwidth]{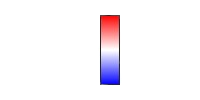}};
\node[anchor=south west,inner sep=0] (image) at (8.45,0.02) {\includegraphics[page=1,trim=35mm 0mm 28mm 0mm, clip,width=0.018\textwidth]{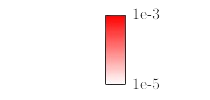}};
\node[anchor=west] at (-0.10,6.60) {$(a)$ flow field};
\node[anchor=west] at (8.10,6.60) {$(b)$ LAD variables};
\scriptsize
\node[anchor=east] at (0.55,2.25) {$u_x$};
\node[anchor=west] at (0.40,2.00) {$2$};
\node[anchor=west] at (0.40,1.35) {$0$};
\node[anchor=east] at (0.45,1.05) {$p$};
\node[anchor=west] at (0.35,0.85) {$1.2$};
\node[anchor=west] at (0.35,0.15) {$0.5$};
\node[anchor=east] at (9.05,1.10) {$\beta^*/\mu$};
\node[anchor=west] at (8.65,0.85) {$1\mathrm{e}{2}$};
\node[anchor=west] at (8.65,0.20) {$1\mathrm{e}{0}$};
\end{tikzpicture} \\ \vspace{0mm}
\caption{Visualizations at the same instant in time of supersonic turbulent flow over a blade at $M_\infty = 2$ and $Re_c = 500,000$ using the sensor $f_{sh}$. $(a)$ isosurfaces of $Q = 100$ colored by $u_x$-velocity, and background contours of $p$. $(b)$ blue isosurfaces of $\mu^*/\mu = 1$ and background slice contours of $\beta^*/\mu$.} 
\label{fig:Hugo2}
\end{figure}
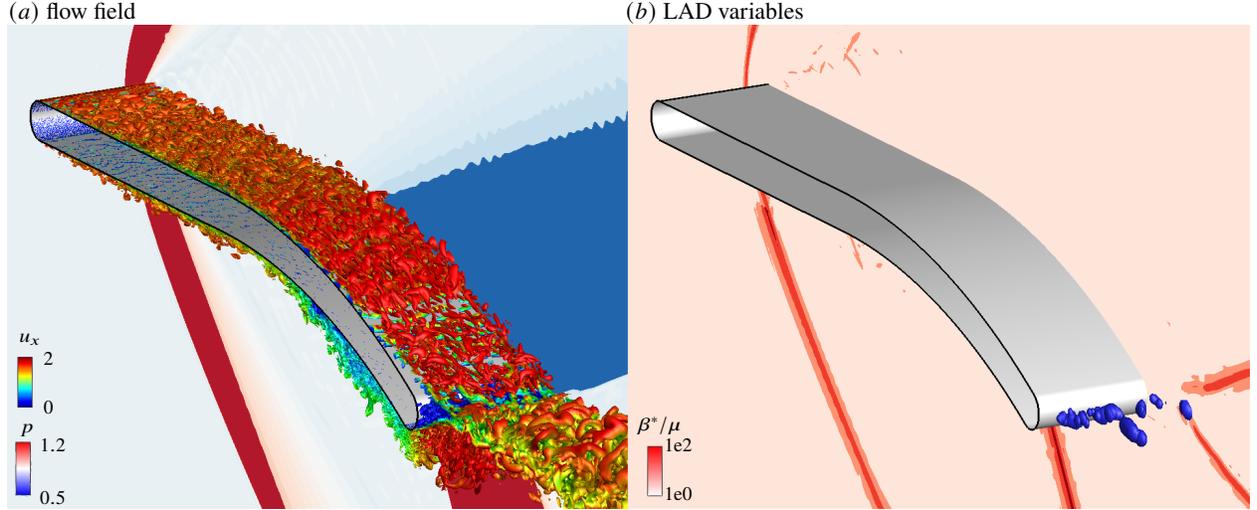

A fully 3D visualization is shown in Fig. \ref{fig:Hugo2}, with the flow field shown on the left in terms of $Q$-criterion colored by $u_x$ velocity with a background plane of pressure contours. In this plane, compression regions appear in red due to the presence of shock waves, while an expansion region appears in blue due to the convex blade curvature. The LAD variables $\beta^*$ and $\mu^*$ are presented on the right. We note that the blue isosurfaces of $\mu^*$ shown in Fig. \ref{fig:Hugo2}$(b)$ emerge near the trailing edge, a critical region due to its significant pressure decay and vorticity generation. In contrast to the standard $\mu^*$ (Fig. \ref{fig:Hugo}(c)), no diffusivity is enforced over the turbulent boundary layer. The addition of the sensor-restrained $\mu^*$ occurs only in the strictly needed regions to secure numerical stability. With a proper sensor $f_{sh}$, we perform such high-fidelity simulation using a high-order numerical scheme maintaining numerical stability of the solver while significantly reducing or fully suppressing the added artificial diffusivity in resolved regions of the flow.

\section{Conclusions}
\label{sec:conclusions}

We propose a flow-field-based Ducros-type sensor to spatially restrain the application of  artificial shear viscosity within the LAD scheme. The proposed approach is designed to avoid high-shear regions such as boundary layers and wakes, minimizing its influence on the resolved turbulent scales. Our approach builds on the standard LAD, where the artificial shear viscosity can be applied in resolved regions of the flow to stabilize simulations while negatively affecting the flow statistics. The proposed sensor-restrained artificial shear viscosity achieves a good compromise between high-resolution of the turbulent scales and numerical stability by suppressing spurious oscillations emerging in low-pressure-core vortices, which are unattended by the artificial bulk viscosity. By testing across transonic and supersonic flows, we demonstrate that this simple sensor effectively stabilizes the flow simulation while applying substantially smaller amounts of artificial shear viscosity to the flow when compared to the standard LAD. In this manner, we prevent an excessive diffusion in the resolved flow scales. This approach has the potential to enable the application of simple LAD models to non-dissipative compact finite-difference schemes to study compressible flows dominated by energetic vortices, which are commonly encountered in separated flows over bluff bodies and wakes.

\section*{Acknowledgments}

J.H.M.~Ribeiro and W.R.~Wolf acknowledge the support from Funda\c{c}\~ao de Amparo \`a Pesquisa do Estado de S\~ao Paulo, FAPESP (grants No.\ 2023/12226-5 and 2021/06448-0). W.R.~Wolf and H.F.S.~Lui also acknowledge the support from the Air Force Office of Scientific Research, AFOSR (grant No.\ FA9550-23-1-0615). We thank the Coaraci Supercomputer for computer time (FAPESP grant No.\ 2019/17874-0) and the Center for Computing in Engineering and Sciences at UNICAMP (FAPESP grant No.\ 2013/08293-7).

\bibliography{biblio}
\bibliographystyle{unsrt} 

\end{document}